%% file: cappellari_CL2010-0283.tex
\documentclass{IEEEtran}

\usepackage[dvips]{graphicx}
\usepackage[tight]{subfigure}
\usepackage[cmex10]{amsmath}
\usepackage{amssymb}

\def \bm {\mathbf}

\begin{document}
\title{On Syndrome Decoding for Slepian-Wolf Coding Based on Convolutional and Turbo Codes}

\author{Lorenzo~Cappellari,~\IEEEmembership{Member,~IEEE}
\thanks{Manuscript received February 17, 2010. The associate editor coordinating the review of this letter and approving it for publication was Prof.~Z.~Yan.}
\thanks{L.~Cappellari is with the Dept.~of Information Engineering, University of Padova, 35131 Padova, Italy (email: lorenzo.cappellari@dei.unipd.it).}}

\maketitle

\begin{abstract}
\input{cappellari_CL2010-0283.txt}
\end{abstract}


\section{Introduction}
Several practical schemes for lossless source coding with side information at the decoder directly embody the optimal binning scheme given by Wyner in \cite{wyner_resultsShannon}. In practice, the syndrome with respect to a turbo (or an LDPC) code that is a \emph{good} channel code for the fictitious \emph{correlation channel} between the source and the side information is used as compressed representation of a given realization. At the decoder the best estimate is found into the \emph{coset} signalled by this syndrome.

In this letter, the focus is on the schemes based on convolutional and (parallel) turbo codes \cite{liveris_DistributedCompressionOf,tu_SF-ISF,tan_EnhancingTheRobustness,roumy_RateAdaptiveTurbo}. All these approaches use in fact a \emph{systematic} syndrome former as source encoder, since this is a computationally efficient solution for syndrome formation w.r.t.~these codes, in particular w.r.t.~turbo codes. Decoding is instead performed with ad-hoc algorithms. Despite this, here it is shown that the resulting estimate is the same in all cases and could be found by simply performing a \emph{maximum a posteriori probability} (MAP) search.

\section{Channel Coding with Convolutional and Turbo Codes}
In channel coding based on an $(n,k)$ convolutional code over $GF(q)$, encoding is often realized by the \emph{systematic} generator $\bm{G}(D)=[\bm{I}_k|\bm{P}(D)]$. The actual structure of the \emph{trellis section} that realizes the $k$-in $(n-k)$-out system $\bm{P}(D)$ is described by the function $\chi_{\sigma_{i-1}}^{\sigma_i}(\bm{x}^s_i,\bm{x}^p_i)$, which indicates if at any time $i$ the transition from state $\sigma_{i-1}$ to state $\sigma_i$ is possible with the \emph{systematic} $k$-tuple $\bm{x}^s_i$ as input and the \emph{parity} $(n-k)$-tuple $\bm{x}^p_i$ as output.

\begin{figure}
\centering
\subfigure[]{
\label{f:cc_factor_graph}
\includegraphics[scale=.66]{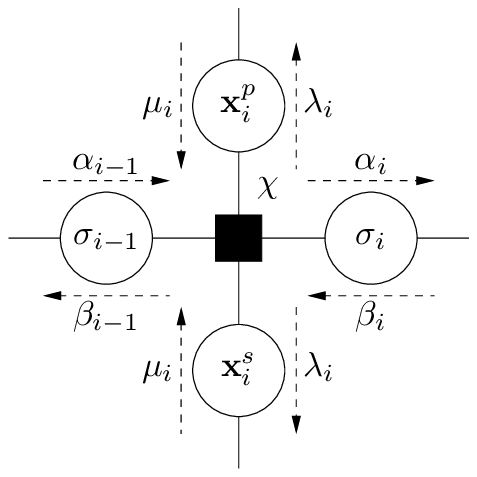}} \hfil
\subfigure[]{
\label{f:sd_factor_graph}
\includegraphics[scale=.66]{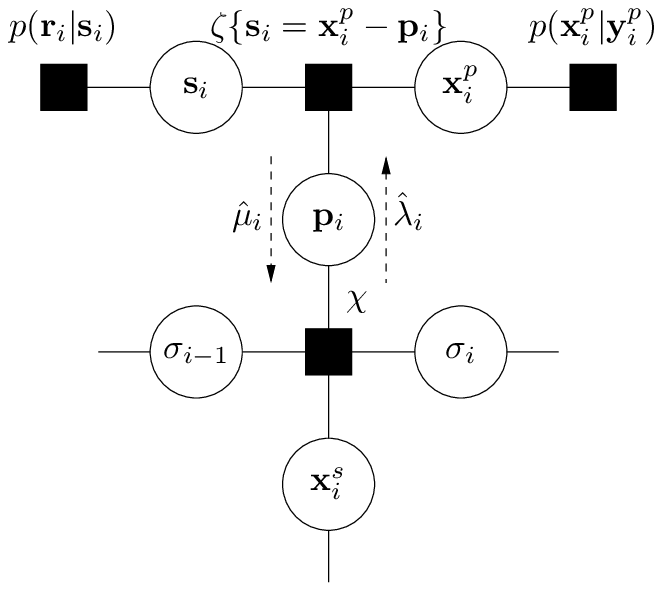}} \hfil
\caption{A portion of the factor-graph used for decoding (a) convolutional codes and (b) systematic syndromes.}
\end{figure}

The symbol-by-symbol MAP decoder evaluates $\arg \max_{\bm{x}^s_i} p\left(\bm{x}^s_i|\bm{y}^s(D),\bm{y}^p(D)\right)$, where $\bm{y}^s(D)$ and $\bm{y}^p(D)$ are the received sequences corresponding to the systematic and to the parity sequence, respectively, sent by the encoder (\emph{formal vector power series} are defined as $\bm{x}(D) \triangleq \sum_i \bm{x}_i D^i$, where $\bm{x}_i$ are row vectors). If the transmission channel is \emph{memoryless}, i.e.~is defined by $p(\bm{y}^s_i|\bm{x}^s_i)$ (and by $p(\bm{y}^p_i|\bm{x}^p_i)$), and the incoming symbols into the encoder are \emph{i.i.d.}, $p(\bm{x}^s_i|\bm{y}^s(D),\bm{y}^p(D))$ can be efficiently found using the BCJR algorithm, that is by means of \emph{message-passing} marginalization over the \emph{factor-graph} shown (according to the convention of \cite{kschischang_FactorGraphsAnd}) in Fig.~\ref{f:cc_factor_graph}. For example, at the $i$-th step of the \emph{forward phase} the input messages $\mu_i(\bm{x}^s_i) = p(\bm{x}^s_i)p(\bm{y}^s_i|\bm{x}^s_i) \propto p(\bm{x}^s_i|\bm{y}^s_i)$ and $\mu_i(\bm{x}^p_i) = p(\bm{y}^p_i|\bm{x}^p_i)$ are used to evaluate the \emph{forward metric}
\begin{equation}
\alpha_i(\sigma_i) = \sum_{\sim\sigma_i} \chi_{\sigma_{i-1}}^{\sigma_i}(\bm{x}^s_i,\bm{x}^p_i) \alpha_{i-1}(\sigma_{i-1}) \mu_i(\bm{x}^s_i) \mu_i(\bm{x}^p_i)\;\nonumber
\end{equation}
($\sum_{\sim x}$ indicates the sum w.r.t.~all involved variables except $x$). Once also the \emph{backward metric} $\beta_i(\sigma_i)$ is available, the optimal choice at time $i$ is found maximizing $\mu_i(\bm{x}^s_i)\lambda_i(\bm{x}^s_i)$, where
\begin{equation}
\lambda_i(\bm{x}^s_i) = \sum_{\sim\bm{x}^s_i} \chi_{\sigma_{i-1}}^{\sigma_i}(\bm{x}^s_i,\bm{x}^p_i) \alpha_{i-1}(\sigma_{i-1}) \beta_i(\sigma_i) \mu_i(\bm{x}^p_i)\;.\nonumber
\end{equation}

In case of turbo coding, two $(n_j,k)$ systematic convolutional encoders are used ($j=0,1$). MAP decoding is approximated with an iterative procedure that alternatively decodes the two received parity sequences. The input messages for decoding one code are taken from the output messages relative to the other one; in particular, $\mu^{(j)}_i(\bm{x}^s_i) \propto p(\bm{x}^s_i|\bm{y}^s_i)\lambda^{(1-j)}_i(\bm{x}^s_i)$.

\section{Syndrome-Based Source Coding}
Given a linear code, any corresponding \emph{parity-check matrix} defines a source encoder as it bins any source sequence according to its \emph{syndrome}. In correspondence of an $(n,k)$ convolutional code, the \emph{systematic} parity-check matrix $\bm{H}(D)=[-\bm{P}^T(D)|\bm{I}_{n-k}]$ represents the most straightforward choice for this purpose. In fact, if the source sequence (of \emph{i.i.d.}~$n$-tuples over $GF(q)$) $\bm{x}(D)$ is broken into the sequences $\bm{x}^s(D)$ and $\bm{x}^p(D)$ composed by $k$- and $(n-k)$-tuples such that $\bm{x}_i = [\bm{x}^s_i|\bm{x}^p_i]$, then the output of the $n$-in $(n-k)$-out encoder $\bm{H}^T(D)$ is $\bm{s}(D) = \bm{x}(D) \bm{H}^T(D) = \bm{x}^p(D) - \bm{p}(D)$, where $\bm{p}(D) = \bm{x}^s(D) \bm{P}(D)$ is the output from $\bm{P}(D)$ with $\bm{x}^s(D)$ as input; with this choice, the source encoder can re-use the same algorithm used in a systematic channel encoder.

In case of turbo codes, up to symbol reordering, the channel encoder realizes a block-based transformation $\bm{G} = [\bm{I}_{Nk}|\bm{P}_0|\bm{P}_1]$ where $\bm{P}_j$ has $N(n_j-k)$ outputs. If the source sequence (of $N(n_0+n_1-k)$ symbols) $\bm{x}$ is suitably broken into the sequences $\bm{x}^s$, $\bm{x}^0$, and $\bm{x}^1$ composed by $Nk$, $N(n_0-k)$, and $N(n_1-k)$ symbols, the systematic \emph{turbo-syndrome} $\bm{s}=[\bm{s}^0|\bm{s}^1]$ can be evaluated as the difference between $[\bm{x}^0|\bm{x}^1]$ and the \emph{turbo-parity} corresponding to $\bm{x}^s$, obtained with a conventional turbo encoder. Each syndrome component $\bm{s}^j$ acts as a systematic syndrome for the corresponding constituent code.

Several algorithms have been proposed in literature for systematic syndrome decoding; if a side information sequence $\bm{y}(D)$ is present, they usually assume that there exists a \emph{memoryless correlation channel} between $\bm{x}(D)$ and $\bm{y}(D)$ defined by $p(\bm{y}_i|\bm{x}_i)$. As shown below, despite they seem different, they actually obtain the exact same source reconstruction; this holds for both convolutional and turbo codes.

\subsection{Principal and Complementary Trellises}
In syndrome decoding, differently from channel decoding, the side information sequence must be decoded over the coset of the original code signalled by the syndrome. In \cite{liveris_DistributedCompressionOf} each other trellis section structure is modified for taking into account this fact. In particular, at time $i$, the actual structure is described by
\begin{equation}
\chi_{\sigma_{i-1}}^{\sigma_i}(\bm{x}^s_i,\bm{x}^p_i;i) \triangleq \chi_{\sigma_{i-1}}^{\sigma_i}(\bm{x}^s_i,\bm{x}^p_i - \bm{s}_i)\;.\nonumber
\end{equation}
In practice, if $\bm{s}_i=0$ the BCJR algorithm uses the \emph{principal} (original) trellis, otherwise it uses one over $q^{n-k}-1$ \emph{complementary} trellises that share the same state transitions of the original one but have different branch labels. Each input message $\mu_i(\bm{x}^p_i)$ is modified too in order to take into account the known prior probability. More precisely, the input messages are $\mu'_i(\bm{x}^s_i) = \mu_i(\bm{x}^s_i)$ and $\mu'_i(\bm{x}^p_i) = p(\bm{x}^p_i)\mu_i(\bm{x}^p_i) \propto p(\bm{x}^p_i|\bm{y}^p_i)$. The forward metrics and output messages are now
{\setlength{\arraycolsep}{0.0em}\begin{eqnarray}
\alpha'_i(\sigma_i) &{}={}& \sum_{\sim\sigma_i} \chi_{\sigma_{i-1}}^{\sigma_i}(\bm{x}^s_i,\bm{x}^p_i;i) \alpha'_{i-1}(\sigma_{i-1}) \mu'_i(\bm{x}^s_i) \mu'_i(\bm{x}^p_i)\nonumber\\
\lambda'_i(\bm{x}^s_i) &{}={}& \sum_{\sim\bm{x}^s_i} \chi_{\sigma_{i-1}}^{\sigma_i}(\bm{x}^s_i,\bm{x}^p_i;i) \alpha'_{i-1}(\sigma_{i-1}) \beta'_i(\sigma_i) \mu'_i(\bm{x}^p_i)\;.\nonumber
\end{eqnarray}}%
The output messages $\lambda'_i(\bm{x}^p_i)$ are also computed to estimate the most probable $\bm{x}^p_i$ as the one maximizing $\mu'_i(\bm{x}^p_i)\lambda'_i(\bm{x}^p_i)$.

\subsection{Inverse Syndrome Formation}
Instead of modifying the trellis structure, in \cite{tu_SF-ISF} a codeword $\bm{c}(D)$ belonging to the signalled coset is formed and subtracted from the side information\footnote{Source and side information must be defined over the same alphabet.} prior to decoding over the original code. The source sequence equals $\bm{x}(D) = \bm{c}(D) + \bm{c}_0(D)$ for some codeword $\bm{c}_0(D)$ belonging to the original code. Since $\bm{y}(D) - \bm{c}(D) = \bm{c}_0(D) + [\bm{y}(D) - \bm{x}(D)]$ and the code is tailored for the correlation channel (assumed linear), $\bm{c}_0(D)$ can be in principle found decoding this difference; eventually, the sum with $\bm{c}(D)$ permits to reconstruct the source sequence.

In practice, in correspondence of the (systematic) syndrome $\bm{s}(D)$ a suitable $\bm{c}(D)$ is the one such that $\bm{c}_i = [\bm{0}|\bm{s}_i]$. Consequently, the actual input messages to the BCJR algorithm are $\mu''_i(\bm{x}^s_i) \propto p(\bm{x}^s_i|\bm{y}^s_i-\bm{0}) \propto \mu'_i(\bm{x}^s_i)$ and $\mu''_i(\bm{x}^p_i) \propto p(\bm{x}^p_i|\bm{y}^p_i-\bm{s}_i)$; if the \emph{backward} correlation channel (i.e.~the one between the side information and the source) is linear, then $\mu''_i(\bm{x}^p_i) \propto p(\bm{x}^p_i+\bm{s}_i|\bm{y}^p_i) \propto \mu'_i(\bm{x}^p_i+\bm{s}_i)$.

Despite the different strategy, the forward and the backward metrics are proportional to the ones evaluated in \cite{liveris_DistributedCompressionOf}\footnote{In this treatment, it is implied that all \emph{initial} metrics $\alpha^\times_0(\sigma_0)$ and $\beta^\times_N(\sigma_N)$ ($N$ is the number of vectors) are the same, apart from some scaling factors.}. For example, in fact, assuming $\alpha''_{i-1}(\sigma_{i-1}) \propto \alpha'_{i-1}(\sigma_{i-1})$,\footnote{As $\bm{x}^p_i$ takes on a finite group, $\sum_{\sim x} f(x, \bm{x}^p_i) = \sum_{\sim x,\bm{s}_i} f(x, \bm{x}^p_i - \bm{s}_i)$.}
{\setlength{\arraycolsep}{0.0em}\begin{eqnarray}
&{}& \alpha''_i(\sigma_i) = \sum_{\sim\sigma_i} \chi_{\sigma_{i-1}}^{\sigma_i}(\bm{x}^s_i,\bm{x}^p_i) \alpha''_{i-1}(\sigma_{i-1}) \mu''_i(\bm{x}^s_i) \mu''_i(\bm{x}^p_i) = \nonumber\\
&{}={}& \sum_{\sim{\genfrac{(}{)}{0pt}{}{\sigma_i}{\bm{s}_i}}} \chi_{\sigma_{i-1}}^{\sigma_i}(\bm{x}^s_i,\bm{x}^p_i-\bm{s}_i) \alpha''_{i-1}(\sigma_{i-1}) \mu''_i(\bm{x}^s_i) \mu''_i(\bm{x}^p_i-\bm{s}_i) \propto \nonumber\\
&{}\propto{}& \sum_{\sim\sigma_i} \chi_{\sigma_{i-1}}^{\sigma_i}(\bm{x}^s_i,\bm{x}^p_i;i) \alpha'_{i-1}(\sigma_{i-1}) \mu'_i(\bm{x}^s_i) \mu'_i(\bm{x}^p_i) = \alpha'_i(\sigma_i)\;.\nonumber
\end{eqnarray}}%
Similarly, the \emph{translated} output messages $\lambda''_i(\bm{x}^s_i-\bm{0})$ and $\lambda''_i(\bm{x}^p_i-\bm{s}_i)$ are proportional to $\lambda'_i(\bm{x}^s_i)$ and $\lambda'_i(\bm{x}^p_i)$, respectively, so that the sum of the MAP estimate that the algorithm finds for $\bm{c}_0(D)$ with $\bm{c}(D)$ coincides exactly with the MAP estimate evaluated in \cite{liveris_DistributedCompressionOf}\footnote{In \cite{tu_SF-ISF}, it is not clear if $\lambda''_i(\bm{x}^p_i)$ is actually maximized or rather a sub-optimal estimate of the \emph{parity} portion $\bm{x}^p(D)$ is found as the output of $\bm{P}(D)$ with the MAP estimate of the \emph{systematic} portion $\bm{x}^s(D)$ as input, in which case the resulting estimate of $\bm{c}_0(D)$ would be a real codeword.\label{n:subopt}}.

\subsection{Decoding Under Parity Perspective}
In \cite{tan_EnhancingTheRobustness}, it is correctly observed that the syndrome received by the decoder is also a \emph{parity} message. In particular, the systematic syndrome corresponds to the parity obtained with the systematic generator $\bm{G}'(D) = [\bm{I}_n|\bm{P}'(D)]$, where $\bm{P}'(D) = \bm{H}(D)^T$. The source is then recovered by jointly decoding the side information and the received parity message with the corresponding channel decoder.

A realization of the system $\bm{P}'(D)$ can be easily derived from a realization of $\bm{P}(D)$. More precisely, its trellis section structure can be described by $\xi_{\sigma_{i-1}}^{\sigma_i}(\bm{x}_i,\bm{s}_i) \triangleq \chi_{\sigma_{i-1}}^{\sigma_i}(\bm{x}^s_i,\bm{x}^p_i-\bm{s}_i)$, where $\bm{x}^s_i$ and $\bm{x}^p_i$ are a $k$-tuple and an $(n-k)$-tuple obtained from $\bm{x}_i$ as above; $\xi_{\sigma_{i-1}}^{\sigma_i}(\bm{x}_i,\bm{s}_i)$ indicates if at any time $i$ the transition from state $\sigma_{i-1}$ to state $\sigma_i$ is possible with the source $n$-tuple $\bm{x}_i$ as input and the syndrome $(n-k)$-tuple $\bm{s}_i$ as output. As a remark, in correspondence of each transition of the original trellis, this trellis has $q^{n-k}$ parallel transitions that share the same sub-label $\bm{x}^s_i$; in practice, this trellis coincides with the \emph{source coding trellis} of \cite{liveris_DistributedCompressionOf}.

The BCJR algorithm takes now $\mu'''_i(\bm{x}_i) = p(\bm{x}_i)p(\bm{y}_i|\bm{x}_i) \propto p(\bm{x}_i|\bm{y}_i) \propto \mu'_i(\bm{x}^s_i) \mu'_i(\bm{x}^p_i) $ and $\mu'''_i(\bm{s}'_i) = p(\bm{r}_i|\bm{s}'_i)$ as inputs, where $p(\bm{r}_i|\bm{s}'_i)$ describes the \emph{memoryless} channel between the syndrome sequence sent and the one \emph{received} $\bm{r}(D)$. Again, if this channel is error free, i.e.~$\mu'''_i(\bm{s}'_i)$ is non zero if and only if $\bm{s}'_i$ equals the actual syndrome $\bm{s}_i$, the forward and the backward metrics involved in decoding equal the ones evaluated in \cite{liveris_DistributedCompressionOf}. For example, in fact, assuming $\alpha'''_{i-1}(\sigma_{i-1}) \propto \alpha'_{i-1}(\sigma_{i-1})$,
{\setlength{\arraycolsep}{0.0em}\begin{eqnarray}
&{}& \alpha'''_i(\sigma_i) = \sum_{\sim\sigma_i} \xi_{\sigma_{i-1}}^{\sigma_i}(\bm{x}_i,\bm{s}'_i) \alpha'''_{i-1}(\sigma_{i-1}) \mu'''_i(\bm{x}_i) \mu'''_i(\bm{s}'_i) \propto \nonumber\\
&{}\propto{}& \sum_{\sim{\genfrac{(}{)}{0pt}{}{\sigma_i}{\bm{s}_i}}} \chi_{\sigma_{i-1}}^{\sigma_i}(\bm{x}^s_i,\bm{x}^p_i-\bm{s}_i) \alpha'''_{i-1}(\sigma_{i-1}) \mu'''_i(\bm{x}_i) \propto \nonumber\\
&{}\propto{}& \sum_{\sim\sigma_i} \chi_{\sigma_{i-1}}^{\sigma_i}(\bm{x}^s_i,\bm{x}^p_i;i) \alpha'_{i-1}(\sigma_{i-1}) \mu'_i(\bm{x}^s_i) \mu'_i(\bm{x}^p_i) = \alpha'_i(\sigma_i)\;.\nonumber
\end{eqnarray}}%
In addition, the MAP estimate for $\bm{x}^s_i$ is found maximizing
$
\sum_{\sim\bm{x}^s_i} \mu'''_i(\bm{x}_i) \lambda'''_i(\bm{x}_i) \propto
\mu'_i(\bm{x}^s_i) \sum_{\sim\bm{x}^s_i} \mu'_i(\bm{x}^p_i) \lambda'''_i(\bm{x}_i)
$, 
where the output message $\lambda'''_i(\bm{x}_i)$ satisfies
{\setlength{\arraycolsep}{0.0em}\begin{eqnarray}
\lambda'''_i(\bm{x}_i) &{}={}& \sum_{\sim\bm{x}_i} \xi_{\sigma_{i-1}}^{\sigma_i}(\bm{x}_i,\bm{s}'_i) \alpha'''_{i-1}(\sigma_{i-1}) \beta'''_i(\sigma_i) \mu'''_i(\bm{s}'_i)\nonumber\\
&{}\propto{}& \sum_{\sim{\genfrac{(}{)}{0pt}{}{[\bm{x}^s_i|\bm{x}^p_i]}{\bm{s}_i}}} \chi_{\sigma_{i-1}}^{\sigma_i}(\bm{x}^s_i,\bm{x}^p_i-\bm{s}_i) \alpha'''_{i-1}(\sigma_{i-1}) \beta'''_i(\sigma_i)\nonumber\\
&{}\propto{}& \sum_{\sim[\bm{x}^s_i|\bm{x}^p_i]} \chi_{\sigma_{i-1}}^{\sigma_i}(\bm{x}^s_i,\bm{x}^p_i;i) \alpha'_{i-1}(\sigma_{i-1}) \beta'_i(\sigma_i)\;.\nonumber
\end{eqnarray}}%
Hence, this MAP estimate maximizes also $\mu'_i(\bm{x}^s_i)\lambda'_i(\bm{x}^s_i)$, i.e.~it is the exact same estimate evaluated in \cite{liveris_DistributedCompressionOf}; the same can be shown for $\bm{x}^p_i$.\footnote{The slight performance increase shown in \cite{tan_EnhancingTheRobustness} w.r.t.~the algorithm in \cite{tu_SF-ISF} can be justified only assuming that a sub-optimal estimate for $\bm{x}^p(D)$ is found in the latter (see note \ref{n:subopt}).}

\subsection{Syndrome Trellis}
In \cite{roumy_RateAdaptiveTurbo} the source-to-syndrome realization implemented by the source encoder is expanded in order to construct a collection of $q^{n-k}$ \emph{syndrome trellises}. In particular, each one of them is constructed using only the encoder state transitions that correspond to the release of a specific syndrome symbol. Then, during decoding, the trellis to be used at time $i$ is specified by the corresponding syndrome symbol received $\bm{s}_i$.

If the source encoder was a systematic syndrome generator, the source-to-syndrome realization could be derived as in the previous section, so that each syndrome trellis would be made of the same state transitions. At time $i$, with  the input message $\mu''''_i(\bm{x}_i)=\mu'''_i(\bm{x}_i) \propto \mu'_i(\bm{x}^s_i) \mu'_i(\bm{x}^p_i)$, the BCJR algorithm would use the trellis described by $\xi_{\sigma_{i-1}}^{\sigma_i}(\bm{x}_i;i) \triangleq \xi_{\sigma_{i-1}}^{\sigma_i}(\bm{x}_i,\bm{s}_i)$. Consequently, the approach would be equal to the one described in the previous section.

However, the fact that the source encoder considered in \cite{roumy_RateAdaptiveTurbo} is \emph{not} a systematic syndrome generator implies a substantial modification to the syndrome decoding process discussed here. Decoding based on the parity perspective as described above can as well handle non-systematic syndromes. A discussion about the advantages offered by this choice is out of the scope of this letter. The first consequence is that, in general, the trellis over which the BCJR algorithm is eventually run no longer shares the trellis transitions used by the systematic channel encoder for the code w.r.t.~which the syndrome is formed.

\section{Maximum a Posteriori Probability Decoding}
The scheme based on inverse syndrome formation is the only one that exactly re-uses the trellis described by $\chi_{\sigma_{i-1}}^{\sigma_i}(\bm{x}^s_i,\bm{x}^p_i)$, but requires that source and side information are defined over the same alphabet. Also, syndromes could be in general corrupted during transmission, but this is not always properly handled by the algorithms described above.

If the \emph{systematic} syndrome is received as $\bm{r}(D)$, the optimal MAP estimate can be simply found as $\arg \max_{\bm{x}_i} p\left(\bm{x}_i|\bm{y}(D)\bm{r}(D)\right)$, i.e.~marginalizing the function $p\left(\bm{x}(D)\bm{p}(D)\bm{s}(D)|\bm{y}(D)\bm{r}(D)\right)$. Apart from some scaling factors, this function is factorable into four terms, given by (i) the backward correlation channel $p\left(\bm{x}(D)|\bm{y}(D)\right)$, (ii) the response $\chi\left(\bm{p}(D)|\bm{x}^s(D)\right)$ of the parity encoder $\bm{P}(D)$, (iii) the response $\zeta\{\bm{s}(D)=\bm{x}^p(D)-\bm{p}(D)\}$ of the systematic syndrome former ($\zeta\{\cdot\}$ indicates the condition in brackets), and (iv) the transmission channel $p\left(\bm{r}(D)|\bm{s}(D)\right)$. If (i) and (iv) are memoryless, once (ii) is exploded into $\sum_{\sim{\genfrac{(}{)}{0pt}{}{\bm{x}^s(D)}{\bm{p}(D)}}}\prod_i\chi_{\sigma_{i-1}}^{\sigma_i}(\bm{x}^s_i,\bm{p}_i)$ the corresponding factor graph is the one shown in Fig.~\ref{f:sd_factor_graph}.

Since the resulting factor graph is a super-graph of the one in Fig.~\ref{f:cc_factor_graph} (and no additional cycles are added), in place of using an ad-hoc solution, syndrome decoding can re-use the same exact algorithm used in channel decoding. In particular, before running the BCJR algorithm, it is sufficient to form the correct input messages $\hat\mu(\bm{p}_i)$ and, after, to properly process the output messages $\hat\lambda(\bm{p}_i)$ in order to form the MAP estimate of $\bm{x}^p_i$. Under this perspective, it does not matter if $\bm{r}(D)$ and $\bm{y}(D)$ are defined over the same alphabet of $\bm{x}(D)$ or not.

\section{Conclusion}
It has been shown that several implementations of systematic syndrome decoding lead actually to the same source reconstruction. The most straightforward and general way to achieve this estimate is to tackle the problem as a MAP one and use iterative message-passing in order to find (or approximate) the optimal solution. This permits to handle \emph{any} correlation and transmission channels without the need of designing a new decoding algorithm.

\bibliographystyle{IEEEtran.bst}
\bibliography{../IEEEabrv,../nonIEEEabrv,../refs_DSC,../refs_other}

\end{document}

%% file: cappellari_CL2010-0283.txt
In source coding, either with or without side information at the decoder, the ultimate performance can be achieved by means of random binning. Structured binning into cosets of performing channel codes has been successfully employed in practical applications. In this letter it is formally shown that various convolutional- and turbo-syndrome decoding algorithms proposed in literature lead in fact to the same estimate. An equivalent implementation is also delineated by directly tackling syndrome decoding as a maximum a posteriori probability problem and solving it by means of iterative message-passing. This solution takes advantage of the exact same structures and algorithms used by the conventional channel decoder for the code according to which the syndrome is formed.